\documentclass[reprint]{revtex4-1}

\date{\today}

\usepackage{graphicx}

\begin{document}

\title{Periodicity of resonant tunneling current induced
by the Stark resonances in semiconductor nanowire}

\author{M. Wo{\l}oszyn}
\email[Electronic address: ]{woloszyn@agh.edu.pl}
\author{J. Adamowski}
\author{P. W\'ojcik}
\author{B.J. Spisak}
\affiliation{AGH University of Science and Technology, Faculty of
Physics and Applied Computer Science, al.~Mickiewicza 30,
30-059~Krakow, Poland}

\begin{abstract}
The modification of the electronic current resulting from Stark resonances
has been studied for the semiconductor nanowire with the double-barrier structure.
Based on the calculated current-voltage characteristics  
we have shown that the resonant tunneling current is a
periodic function of the width of the spacer layer. We have also
demonstrated that the simultaneous change of the source-drain voltage and
the voltage applied to the gate located near the nanowire leads to
almost periodic changes of the resonant tunneling current as a
function of the source-drain and gate voltages.  The periodic
properties of the resonant tunneling current result from the formation
of the Stark resonance states.  If we change the electric field acting in
the nanowire, the Stark states periodically acquire the
energies from the transport window and enhance the tunneling current
in a periodic manner.  We have found that the separations
between the resonant current peaks on the source-drain voltage scale
can be described by a slowly increasing linear function
of the Stark state quantum number.  This allows us
to identify the quantum states that are responsible
for the enhancement of the resonant tunneling.
We have proposed a method of the experimental
observation of the Stark resonances in semiconductor double-barrier
heterostructures.
\end{abstract}


\maketitle

\section{Introduction}\label{sec:intro}

The Wannier-Stark resonance states have been attracting substantial interest
for many years.\cite{Wannier1962,Wilkinson1996,Morifuji1997,Gluck2002,Rosam2003}
These states are created if the electron is simultaneously
subjected to a periodic field, e.g., of the crystal or the
superlattice, and to a uniform electric field.\cite{Gluck2002,bookSchafer2002}
The joint effect of these fields leads to the formation
of the Wannier-Stark states that exhibit the double
periodicity, namely, the periodic localization in the real space
and the periodic sequence of energy levels.\cite{Gluck2002}
The Wannier-Stark states have been observed experimentally in
semiconductor superlattices by measuring the Zener tunneling
current.\cite{Morifuji1997} The lifetime of the Wannier-Stark
resonances has been studied both experimentally and theoretically by
Rosam et al.\cite{Rosam2003} The Wannier-Stark resonances have
been also observed in the optical lattices, e.g., for the system
of cold atoms in the accelerated standing laser wave.\cite{Wilkinson1996}
Recently, the essential progress has been reported
both in the experimental\cite{Sobolev2010,Lima2010,Tackmann2011}
and theoretical\cite{Kolovsky2013} studies of these states,

Similar resonance states can be formed if the uniform electric field is
applied to a semiconductor quantum-well heterostructure.\cite{Austin1985,Ahn1986,Austin1988}
Since the periodic field is not necessary to the formation of these states, 
as it was demonstrated for isolated quantum wells\cite{Austin1985,Ahn1986}
and aperiodic structures\cite{Spisak2009PRB}, they are called the Stark resonances.
The Stark resonances can also occur in the double-barrier resonant tunneling
structures under an external electric field.\cite{Peng1991,Niculescu2000}
The effect of the electric field on the energy of the Stark resonances and on the
resonance lifetimes has been studied theoretically for a single quantum
well\cite{Austin1985,Ahn1986,Borondo1986} and for double-barrier
structures.\cite{Peng1991,Porto1994,Bylicki1996,Niculescu2000,Zambrano2002}
One can expect that the Stark resonances will be observed in the measurements of
the current-voltage characteristics of the devices containing
the double-barrier structures, which can be fabricated  either
in mesa-type layer structures or in nanowires.\cite{Bjork2002}

The electronic transport in  the double-barrier structures embedded in nanowires
has been studied experimentally in
Refs.~\onlinecite{Bjork2002,Fuhrer2007APL,Fuhrer2007NL,Pfund2007}.
The theory of the coherent electronic transport in nanowires with Gaussian-type
scatterers has been elaborated in the
papers.\cite{Bardarson2004,Gudmundsson2008}
The modification of the electron transport in the nanowire due to the
quantum ring, dot, and barrier has been studied by
Gudmundsson et al.\cite{Gudmundsson2005}
In Ref.~\onlinecite{Sowa2010}, the current-voltage characteristics of the
nanowire with the embedded double-barrier structure have been determined for
different spacer widths.
It has been shown\cite{Sowa2010} that the resonant current peak possesses the
asymmetric Fano-type shape for the sufficiently narrow spacers.

In the present paper, we focus on the effect of the spacer width and
the source-drain and gate voltages on the resonant tunneling current through a semiconductor
nanowire containing a double-barrier structure.  We demonstrate that
the resonant current peaks periodically change their heights
if we change the spacer width and/or source-drain voltage
with the suitably tuned gate voltage.
This periodicity results from the formation of the Stark resonance states.

The paper is organized as follows:  the theoretical model is described in
Sec.~\ref{sec:theory}, the results of the calculations are presented in
Sec.~\ref{sec:result} and discussed in Sec.~\ref{sec:disc}.
Section~\ref{sec:concl} contains conclusions and summary.

\section{Theory}\label{sec:theory}

\begin{figure}
\centering
\includegraphics{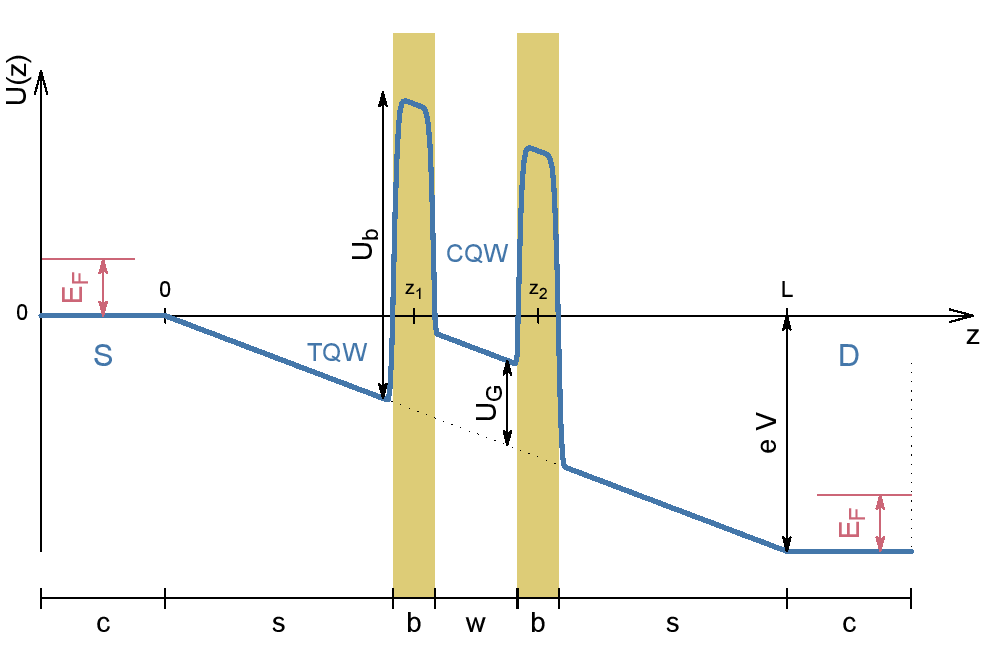} % Model
\caption{(color online)
Potential energy profile of the double-barrier structure
embedded in the nanowire.
Voltage $V$ is applied between the source (S) and drain (D) contacts,
CQW denotes the central quantum well, TQW denotes the triangular quantum well
formed in the left spacer region.
The CQW potential energy bottom is shifted by $U_G=\alpha V_G$ if gate
voltage $V_G$ is applied to the side gate.  $E_F$ is the Fermi energy,
$L$ is the length of the nanowire, i.e., the source-drain distance, the barriers possess height $U_b$ and width
$b$, $w$ is the width of CQW, $s$ is the width of each spacer layer, $c$ is the
width of each contact, and $z_1,z_2$ are the coordinates of the barrier centers.}
\label{fig:model}
\end{figure}

We study the electron transport in the semiconductor nanowire with the
double-barrier structure (Fig.~\ref{fig:model}).
In an external uniform electric field acting parallel to the nanowire axis, the
electronic current can be described within the effective mass approximation by
the one-dimensional (1D) model.
We consider the circuit with the source (S) and drain (D) contacts attached to
both the ends of the nanowire.
Voltages $V_S$ and $V_D$ applied to these contacts result in the source-drain
voltage $V = V_S-V_D > 0$ that generates the uniform electric field
$\mathbf{F} = (0,0,F)$, where $F=-V/L$ and
$L$ is the the length of the nanowire, i.e., the source-drain distance.
The source and drain contacts are made from the heavily doped semiconductors.
Throughout the present paper, the energy of the conduction band bottom of the
source is taken as the reference energy and put equal to zero, moreover, we take
on $V_S=0$.
In the calculations, we also include the effect of the side gate electrode
assumed to be a ring surrounding the nanowire in the region of the central
quantum  well (CQW).
If voltage $V_G$ is applied to the gate, the  potential energy  of the
bottom of the CQW is changed by $U_G=\alpha V_G$, where $\alpha$ is the
voltage-to-energy conversion factor.\cite{bookAdamowski2006}

We consider the nanowire that consists of the CQW with width $w$ surrounded by
the two barriers of equal width $b$ that are separated from the contacts by the
spacers of the same width $s$, i.e., $L=w+2b+2s$.
The potential energy (Fig.~\ref{fig:model})
\begin{equation}
U(z) = U_{db}(z) + U_F(z)
\label{eq:U}
\end{equation}
is the sum of the double-barrier potential energy $U_{db}(z)$ and the potential
energy $U_F(z)$ of the electron in electric field $\mathbf{F}$ of finite range $L$, where
\begin{equation}
U_F(z) = \left\{
\begin{array}{ll}
0 & \mbox{for $z \leq 0 $} \;, \\
e F z & \mbox{for $0 \leq z \leq L $} \;, \\
-e V & \mbox{for $z \geq L$} \;.
\end{array}
\right.
\label{eq:UF}
\end{equation}
The double-barrier potential energy is taken on in the form of the two-center
power-exponential function\cite{Ciurla2002,Kwasniowski2008}
\begin{equation}
U_{db}(z) = U_b \left[
\mathrm{e}^{ - \left( \frac{z-z_1}{R}\right)^p } +
\mathrm{e}^{ - \left( \frac{z-z_2}{R}\right)^p }
\right] \;,
\label{eq:UB}
\end{equation}
where the height of the barriers, $U_b$, is measured with respect to the energy
of the nearby regions of the spacers.
Parameter $p$ describes the sharpness  of the interfaces: for $p=2$ we obtain
the soft Gaussian potentials, while for $p\to\infty$ we get the two rectangular
barriers of width $b=2R$ centered at $z_1=s+b/2$ and $z_2 = z_1 + w + b/2$.

Electronic current $I$ is calculated within the Landauer formalism using the
relation\cite{bookDiVentra2008}
\begin{equation}
 I(V) = \frac{2 e}{h} \int\limits_{-\infty}^{\infty} T(E,V)
        \left[ f_{S}(E) - f_{D}(E) \right] \mathrm{d}E \;,
 \label{eq:IT}
\end{equation}
where $T(E,V)$ is the transmission coefficient.
The electrons in the contacts are described by the Fermi-Dirac distribution
function
\begin{equation}
f_{S,D}(E) = \frac{1}{1+\exp\left( \frac{E-\mu_{S,D}}{k_{B}T} \right) } \;,
\label{eq:fFD}
\end{equation}
where $\mu_{S}$ and $\mu_{D}$ are the electrochemical potentials of
the source and drain, respectively.
We calculate $T(E,V)$ by the transfer matrix method with potential energy
(\ref{eq:U}) approximated by a piecewise constant function.

In order to focus on the effects of geometry and external voltages on the
electronic current, we have performed the calculations for $T=0$ and neglected
the electron scattering.
At zero temperature, the distribution functions (\ref{eq:fFD}) become the simple
step functions with the electrochemical potentials $\mu_S=E_F$ and
$\mu_D=E_F-eV$, where the Fermi energy, $E_F$, is assumed to be the same
for the source and drain.
In this case, the integration in Eq.~(\ref{eq:IT}) runs over the transport
window of width $E_F$ and can be performed numerically.

The numerical calculations have been performed for the InAs
nanowire with the InP barriers\cite{Bjork2002} for the fixed
values of the following material parameters: $U_b=0.6$~eV, $b =
5$~nm (hence $R=2.5$~nm), $w=15$~nm, and $E_F=2$~meV. Because the
widths of the InP barriers are much smaller than length of the
nanowire, we assume that the electrons are described by the InAs
conduction-band mass, i.e., we take on $m=0.0265 m_0$, where $m_0$
is the free-electron rest mass. Putting $p=$~6 we account for the
observed non-perfect sharpness of the InAs/InP
interfaces.\cite{Bjork2002,Niquet2008,Thelander2004}  However, we
have found that the calculated current-voltage characteristics are
rather insensitive to the variation of $p$ provided that the
potential-energy profile is sufficiently steep at the interface.
Therefore, the periodic behavior of the resonant current peaks
presented in Sec.~\ref{sec:result} is the same for $p \rightarrow
\infty$, i.e., for the rectangular potential barriers and well.

We have calculated the current-voltage characteristics using Eq.~(\ref{eq:IT})
for different spacer widths $s$, i.e., for the nanowires with different lengths
$L$.
Due to the geometric symmetry of the nanodevice the change of the spacer width
by $\Delta s$ corresponds to the change of the length of the nanowire by
$\Delta L = 2 \Delta s$.
In order to carry out the transfer-matrix calculations, we have introduced the
$z$-coordinate mesh with $N_L$ mesh points
in the interval $0 \leq z \leq L$.
$N_L$ has been chosen in such a manner that the minimal distance between the nearest mesh
points on the $z$-axis is equal to $0.33$~nm, which gives, e.g., $N_L=6000$ for
$L=2000$~nm.

\section{Results}\label{sec:result}

\subsection*{A. Periodicity of resonant current peaks}

\begin{figure}%[ht]
\centering
\includegraphics{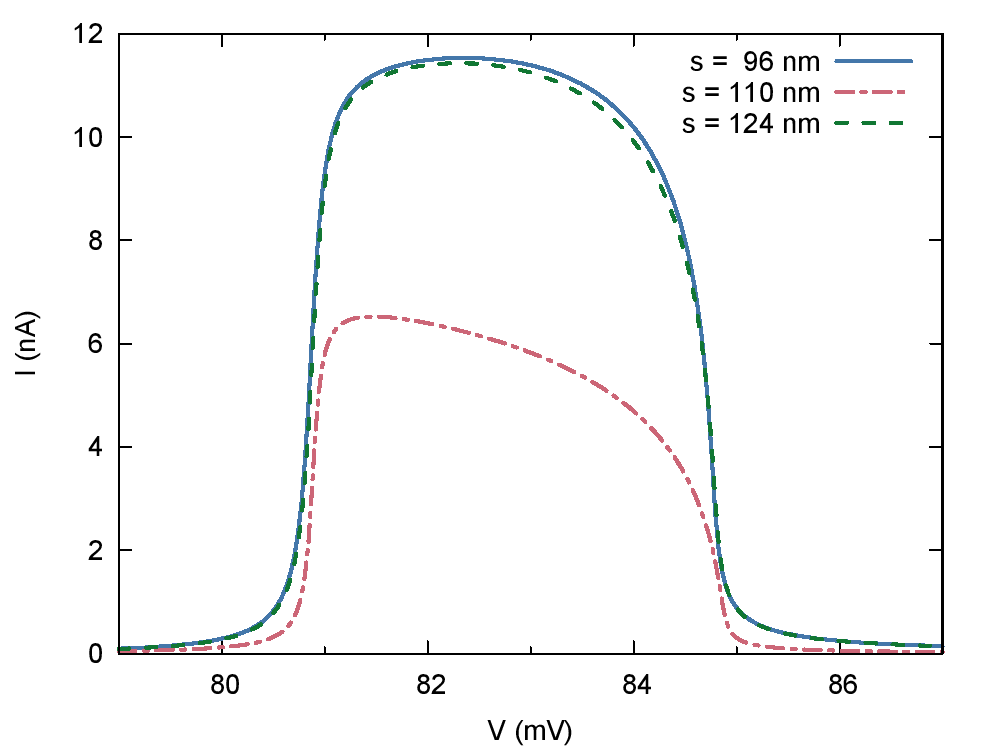} % IV
\caption{(color online) Current-voltage characteristics of the nanowire with the
double-barrier structure calculated for spacer width $s$ = 96 nm (solid, blue
curve), 110 nm (dash-dotted, red curve), and 124 nm (dashed, green curve).}
\label{fig:IV}
\end{figure}

\begin{figure}
\centering
\includegraphics{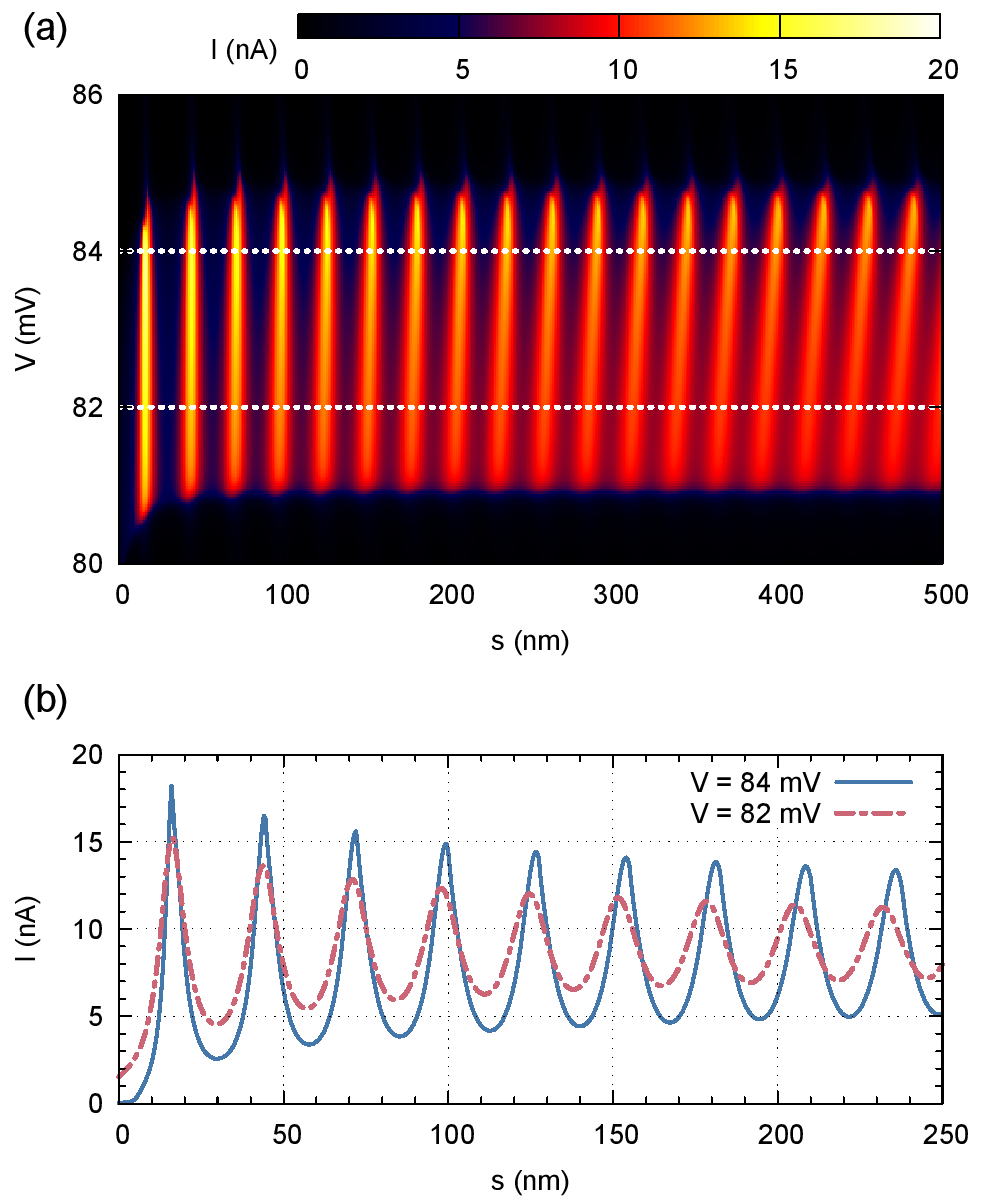} % ILV
\caption{(color online)
(a) Current $I$ as a function of source-drain voltage $V$ and spacer
width $s$.
The dotted lines are drawn for $V = 82$~mV and $V= 84$~mV.
(b) Oscillations of the resonant tunneling current $I$ for $V=82$~mV and $V=84$~mV
as a function of $s$.
Panels (a) and (b) show the results for $V_G=0$.}
\label{fig:IL}
\end{figure}

The present calculations are based on the following physical background: at zero
temperature the incident electron with energy $E_{in}$ can tunnel from the
source to the drain through the double-barrier heterostructure in a resonant
tunneling process if energy $E_{in}$ is aligned with energy $E_{CQW}^{\nu}$ of
quasi-bound state $|\nu\rangle$ localized in the CQW and both the energies
fall into the transport window, i.e.,
\begin{equation}
 0 \leq  E_{in} \simeq E_{CQW}^{\nu} \leq E_F \;.
\label{eq:res_cond}
\end{equation}
If condition (\ref{eq:res_cond}) is satisfied, then -- for suitably chosen
spacer width -- we obtain the strong resonant tunneling current peak on the
current-voltage characteristics that takes on  the shape typical for the
resonant tunneling diode\cite{Sowa2010} (solid, blue curve in Fig.~\ref{fig:IV}).
However, if we perform the calculations  for the spacer that is wider by
$\Delta s =14$ nm, the current peak becomes weak, i.e., its
height is smaller by a factor of two in comparison to the height of the strong
peak.
After increasing the spacer width by $2 \Delta s$, the current-voltage
characteristics with the strong resonant peak is recovered (Fig.~\ref{fig:IV}).
We have found that the similar sequence of the strong and weak resonant current
peaks is repeated every time, if we increase or decrease the spacer width by the
same period of approximately 28~nm.
The periodic changes of the resonant current peaks are presented in
Fig.~\ref{fig:IL}.

Fig.~\ref{fig:IL}(a) displays the resonant current peak that spreads
from $V \simeq 81$~mV to $V \simeq 85$~mV on the current-voltage characteristics.
This peak results from  the resonant tunneling  via the lowest-energy level
$E_{CQW}^{(1)} = 41$~meV that satisfies the resonant-tunneling condition
(\ref{eq:res_cond}).
If we change the spacer width, i.e., the source-drain separation,
the resonant tunneling current changes considerably.
Fig.~\ref{fig:IL}(b) shows that the resonant tunneling current peaks change periodically
with the spacer width.
Based on these results we have estimated that the subsequent strong/weak
resonant current peaks are separated by $P_s \simeq 28$~nm.
Fig.~\ref{fig:IL}(b) also shows that -- for $V = 84$~mV -- the strong resonant
current peak is 3-5 times higher than the weak resonant current peak.
The largest difference between the strong and weak peaks can be
observed for the narrow spacers, i.e., for $s \leq 50$~nm.

\begin{figure}
\centering
\includegraphics{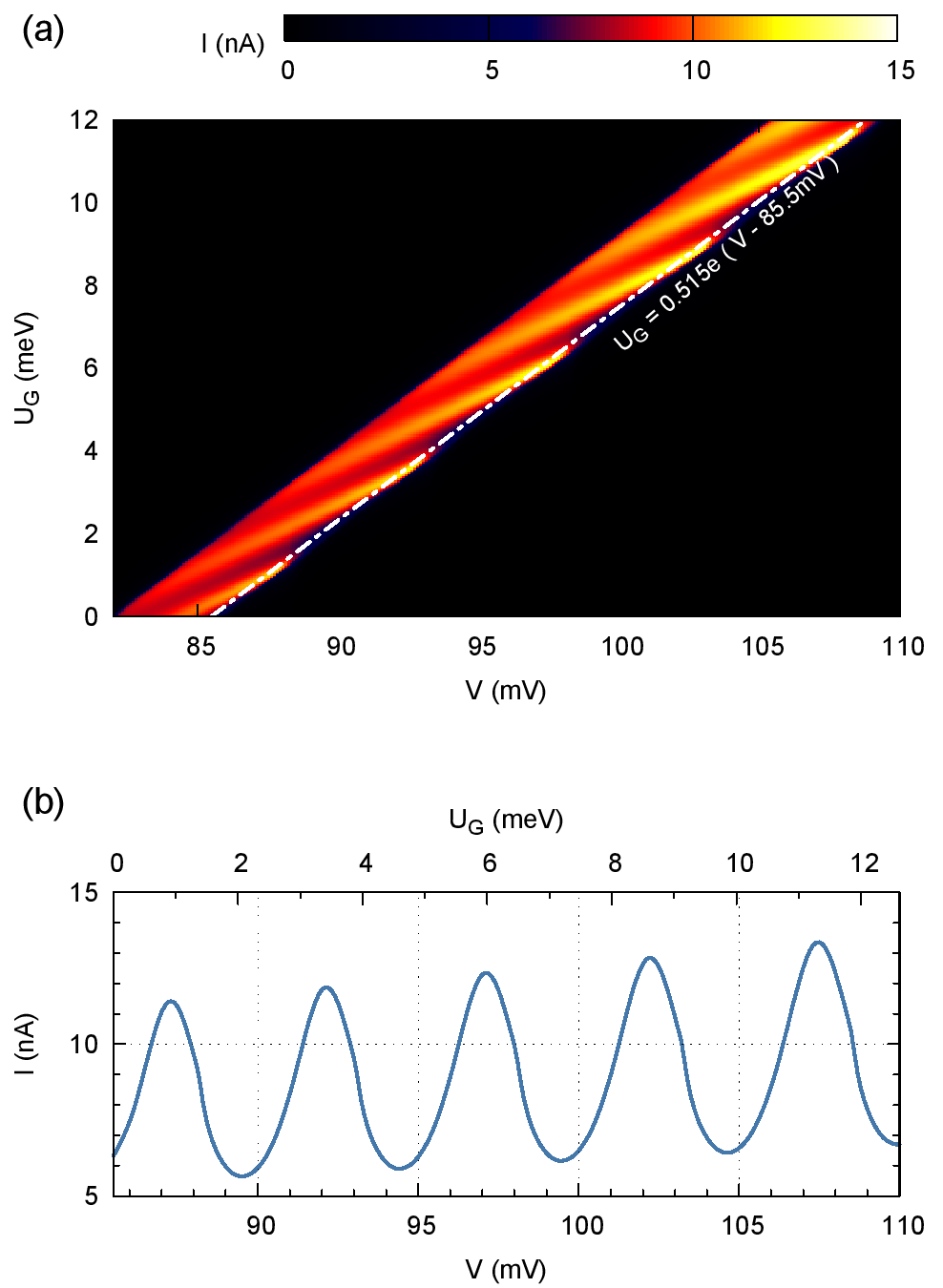} % IUV
\caption{(color online)
(a)~Current $I$ as a function of the source-drain voltage $V$ and the
gate-voltage induced change of the CQW potential energy $U_G$ for a constant
spacer width $s=1000$~nm and for $w=15$ nm.
Gate voltage $V_G$ can be extracted from $U_G$ using the formula
$V_G = U_G/\alpha$.
(b) Oscillations of the resonant tunneling current $I$ obtained for $V$ varying along
the dash-dotted line on panel (a).}
\label{fig:IU}
\end{figure}

There appears a question if can we modify the periodicity of the resonant current
flowing through the nanowire by applying the additional voltage to the side gate?
In order to answer this question, we have simulated the effect
of gate voltage $V_G$ by shifting the CQW potential energy bottom by
$U_G = \alpha V_G$ (cf. Fig.~\ref{fig:model}).
Since the actual value of $\alpha$ depends on the parameters of
the nanostructure and also on the gate voltage,\cite{bookAdamowski2006}
we have used in the calculations -- instead of the gate voltage --
the suitably chosen values of potential energy shift $U_G$.
The results are depicted in Fig. \ref{fig:IU}.
For the negative gate voltage and for $\alpha < 0$ the potential energy changes by
$U_G >0$, which shifts upwards the energy levels of the quasi-bound states localized
in the CQW.  This in turn causes that the resonant tunneling condition
(\ref{eq:res_cond}) is no longer satisfied, i.e., the resonant tunneling is broken.
Therefore, in order to recover the resonant tunneling we have to change the
source-drain voltage $V$ in an appropriate manner.
We have found that changing simultaneously the voltages $V$ and $V_G$ according
to the relation
\begin{equation}
\alpha V_G=0.515 e (V-V_0) \;,
\label{eq:VUG}
\end{equation}
where $V_0 = 85.5$~mV, allows us to follow the maxima of the resonant
tunneling current and its periodic changes (Fig.~\ref{fig:IU}).
Prefactor 0.515 is the lever factor for the source-drain voltage.
Its value (slightly larger than the value 0.5, which is appropriate for the
ideally symmetric system) accounts for a slight asymmetry of the nanodevice
due to the electric field acting between the source and the drain.
If the gate voltage is changed in such a way that
$U_G$ increases from 0 to 10~meV, the position of the resonant current peak
on the current-voltage characteristics is shifted from $V \simeq 85$~mV to
$V \simeq 105$~mV (Fig.~\ref{fig:IU}).
The corresponding sequence of the resonant current peaks is visible along the
stripe on Fig.~\ref{fig:IU}(a).
Fig.~\ref{fig:IU}(b) shows that the height of the resonant tunneling current peak
changes by a factor of two if the source-drain and gate voltages are varied
according to Eq.~(\ref{eq:VUG}), i.e.,
along the straight line displayed on Fig.~\ref{fig:IU}(a).
We have estimated the average separation between the neighboring strong
peaks to be $P_V \simeq 5.1$~mV,
which corresponds to the change of the CQW potential energy by $\sim 2.6$~meV.
We note that the almost periodic changes of the height of the resonant current
peak depicted in Fig.~\ref{fig:IU} have been obtained for the fixed geometric
parameters of the nanodevice, in this case, for $s=1000$~nm and $w=15$~nm.

However, after a closer inspection of the results shown in Fig.~\ref{fig:IU}(b)
we have found that -- on the contrary to the constant separation between the strong
resonant tunneling current peaks as a function of the spacer width
[Fig.~\ref{fig:IL}(b)] -- the separations between the current peaks
[Fig.~\ref{fig:IU}(b)] slightly increase with the increasing source-drain voltage.
This effect will be explained in Section \ref{sec:disc}.

\begin{figure}
\centering
\includegraphics{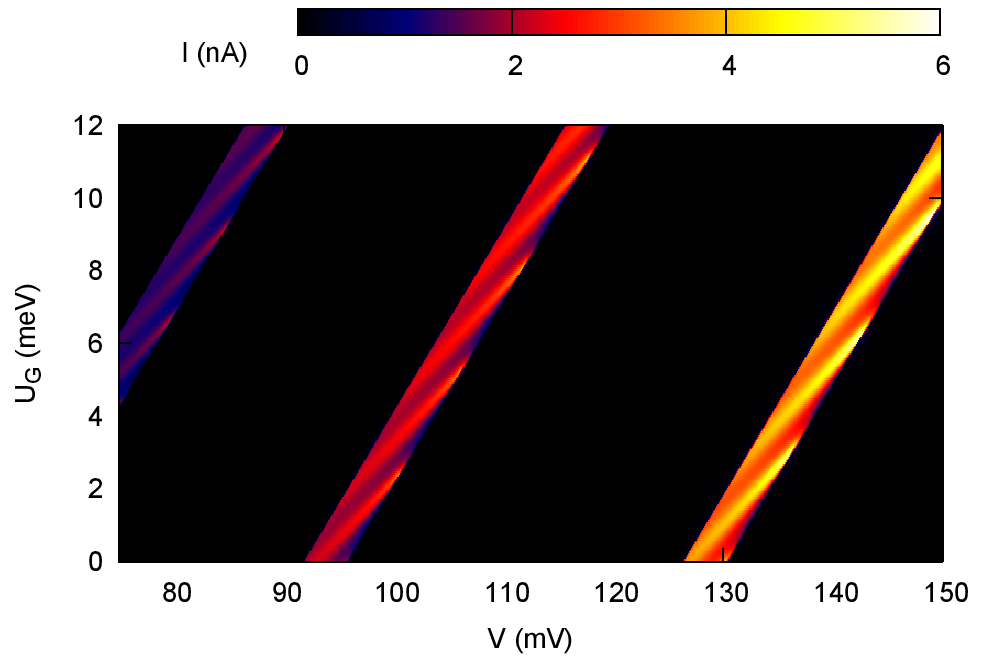} % IUVw
\caption{(color online)
Current $I$ as a function of the source-drain voltage $V$ and potential energy
$U_G$ for $w=100$~nm and $s=1000$~nm.}
\label{fig:IUVw}
\end{figure}

The width of the side gate, to which voltage $V_G$ is applied, should be
approximately equal to width $w$ of the CQW.
In the recently fabricated nanodevices,\cite{Tomioka2012} the ring-shaped gates
that surround the nanowires are much wider than that used to obtain the results
presented in Fig.~\ref{fig:IU}.
In order to check if the periodic behavior of the resonant current peaks also
occurs for the wider gates and CQW's, we have performed the calculations for
$w=100$~nm.
The results are displayed in Fig.~\ref{fig:IUVw}.
We have found that the periodic sequence of the strong and weak resonant current
peaks is still visible when the gate voltage is changed with the increasing bias voltage
according to formula (\ref{eq:VUG}).
As opposite to the narrow CQW (Fig.~\ref{fig:IU}), we obtain more than one
resonant current peak within the same range of the source-drain voltage, which
results from the fact that for the wider CQW energy levels $E_{CQW}^{\nu}$
possess the lower energies and the energy separations between them are smaller.
Therefore, more than one CQW energy level enters the transport window in the considered
source-drain voltage range.
In Fig.~\ref{fig:IUVw}, this effect is shown as the three stripes that
correspond to the three states with subsequent energy levels $E_{CQW}^{\nu}$
The direct numerical calculation of energy levels $E_{CQW}^{\nu}$ has allowed us
to identify these quantum states as described by quantum numbers $\nu = 5,6,7$.

\subsection*{B. Stark resonances}

\begin{figure}
\centering
\includegraphics{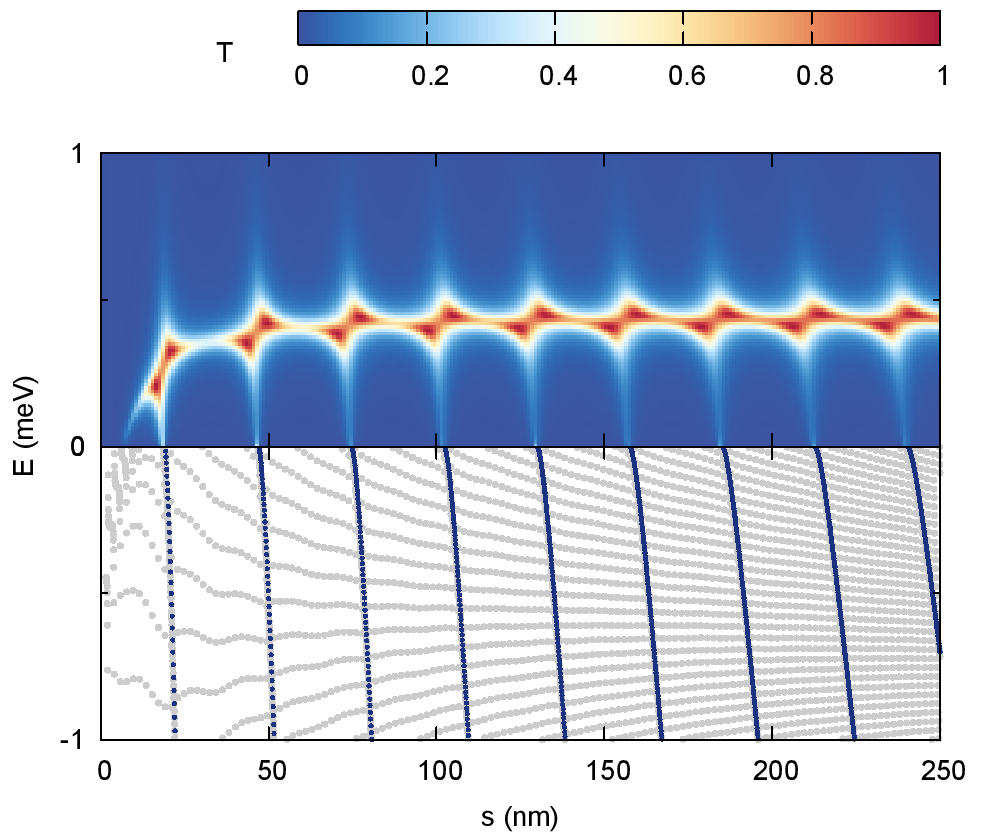} % TLE
\caption{(color online)
Upper part: Transmission coefficient $T$ as a function of energy $E$ and
spacer width $s$.
Lower part: energy levels $E_n$
of the quasi-bound states as functions of the spacer width $s$.
Gray dots show the energies calculated for the entire nanodevice (including the
contacts) and blue dots show the energy levels of the quasi-bound states
localized in the left spacer.
The calculations have been performed for $V=84$~mV and $U_G=0$.}
\label{fig:TLE}
\end{figure}

\begin{figure}
\centering
\includegraphics{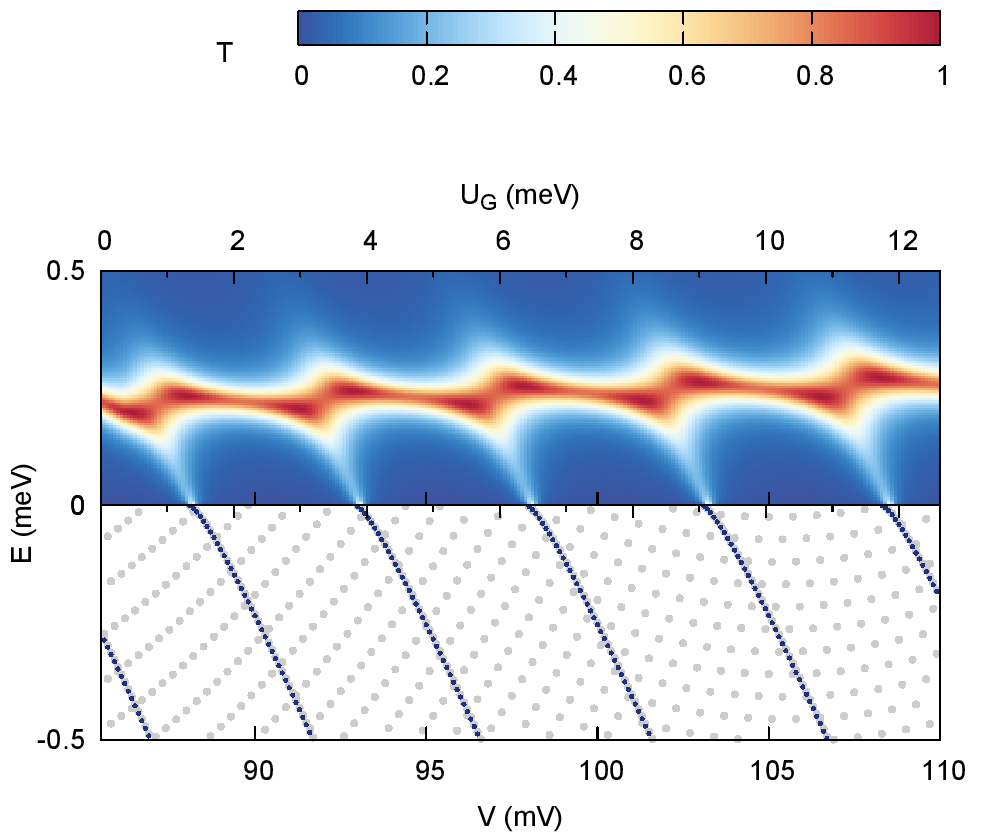} % TUE
\caption{(color online)
Upper part: Transmission coefficient $T$ as a function of energy $E$ and
source-drain voltage $V$ that varies with the gate voltage according to
Eq.~(\ref{eq:VUG}).
Lower part: energy levels $E_n$ of the quasi-bound as functions of the
source-drain voltage $V$ varying with the gate voltage according
to Eq.~(\ref{eq:VUG}).
Gray dots show the energies calculated for the entire nanodevice (including the
contacts) and blue dots show the energy levels of the quasi-bound states
localized in the left spacer.
The calculations have been performed for  $s=1000$~nm and
$U_{G}=0.515 e (V-85.5\,\textrm{mV})$.}
\label{fig:TUE}
\end{figure}

In order to find the physical interpretation of the periodic behavior of the
resonant tunneling current, we have analyzed the transmission coefficient as a
function of the spacer width and the source-drain voltage.
The corresponding plots are presented in the upper parts of Figs.~\ref{fig:TLE}
and \ref{fig:TUE}.
In the lower parts of these figures, we plot energy levels $E_n < 0$ of
quasi-bound states $|n\rangle$ localized in the triangular quantum well
(TQW) that is formed in the left spacer due to
the uniform external electric field acting parallel to the nanowire axis.
Energy levels $E_n < 0$  have been calculated by the second-order
finite-difference method by diagonalizing the resulting Hamiltonian matrix in
the computational box expanded by $c=\pm 3L$ to the right and to the left from
the right and left spacer boundaries, respectively (cf. Fig.~\ref{fig:model}).
In this way, we have checked that the physically relevant results do not depend
on the size of the computational box.
In Figs. \ref{fig:TLE} and \ref{fig:TUE}, the gray and blue dots correspond to
energy levels $E_n$ calculated with the expanded computational box.
The majority of the quantum states associated with these energy levels are the
states localized in the right contact.
The energy levels associated with these states are marked by the gray dots
in the lower parts of Figs.~\ref{fig:TLE} and \ref{fig:TUE}
[see also Fig.~\ref{fig:TR}(b)].
These states are the eigenstates of the system with the potential-energy profile,
which includes the flat potential energy regions corresponding to the contacts
(cf. Fig.~\ref{fig:model}),
and result from the binding of the electron in the computational box of finite size,
i.e., they are unphysical.  However, the fact that we have obtained
these states proves the reliability of the finite difference method applied
and means that we have not omitted any eigenstates of the system in the calculations.
The physical meaning can be attributed to the states with the energy levels
depicted by the blue dots in Figs.~\ref{fig:TLE} and \ref{fig:TUE},
which form almost vertical lines in Fig.~\ref{fig:TLE}.
The energy levels plotted by the blue dotted curves in Figs.~\ref{fig:TLE},
\ref{fig:TUE}, and \ref{fig:TR}(b) have been calculated for the simplified shape
of the potential energy shown by the blue solid lines in Fig.~\ref{fig:TR}(a).
It is interesting that these energy levels exactly agree with those calculated
for the entire system with the double-barrier potential (the gray dots along the
blue curves are not visible since they coincide with each other).
The results of the calculations with the potential energy represented by only
the left part of the potential-energy profile allow us to interpret the
physically relevant energy levels $E_n$ as associated with the quasi-bound
states $|n\rangle$ localized in the TQW in the left spacer region.

We see that both the transmission coefficient $T$ and the energy levels $E_n$
are periodic functions of spacer width $s$ for fixed $V$ (Fig.~\ref{fig:TLE}) or
the source-drain voltage and gate voltage for fixed $s$ (Fig.~\ref{fig:TUE}).
Using Eq.~(\ref{eq:IT}) we can state
that the periodicity of the transmission coefficient leads to the periodicity
of the resonant tunneling current.
The estimated average values of periods $P_s$ and $P_V$
almost exactly coincide with the periods of
the strong/weak resonant current peak sequence estimated from Figs.~\ref{fig:IL}
and \ref{fig:IU}.
For $U_G=0$ (Fig.~\ref{fig:TLE}) the subsequent maxima of the transmission
coefficient appear for the energy of the incident electrons
$E_{in} \simeq 0.45$~meV.
This energy is aligned with the lowest-energy level of the quasi-bound state
localized in the CQW for $V=84$~mV.
Along the straight line $E=E_{in}$ the transmission coefficient as a function of
spacer width $s$ periodically reaches the value $T=1$, which demonstrates that
we deal with the resonant tunneling effect with the periodic changes of the transmission.
Only for the narrow spacers with $s<50$~nm the maximum of transmission
coefficient becomes smaller than 1 and is shifted towards the lower energy
values.\cite{Sowa2010}

If the spacer width decreases, the (negative) energy of the quasi-bound state
increases reaching $E=0$ and continuously goes over into the curve corresponding
to the energy of the resonance state with $E_n>0$.
The transmission maxima  correspond to the resonance states with energies $E>0$
and finite widths that are clearly visible on Fig.~\ref{fig:TLE}.

Of course, the same mechanism is responsible for the creation of the resonances when the
geometric sizes of the device are fixed, but the source-drain voltage is changed
together with the gate voltage according to formula (\ref{eq:VUG}).
Fig.~\ref{fig:TUE} displays the outcome of the similar calculations as in
Fig.~\ref{fig:TLE}, but now the source-drain voltage is altered at fixed $s$, which shifts
the energy levels of the quasi-bound states localized in the CQW.
Therefore, in order to satisfy the resonant tunneling  condition we have to
compensate this shift by applying the suitably chosen gate voltage.
The increasing  source-drain voltage
causes that the depth of the TQW generated in the left spacer also increases,
which leads to the growth of the number of the quasi-bound states localized in
the TQW.

\section{Discussion}\label{sec:disc}

Based on the results presented in Figs.~\ref{fig:TLE} and \ref{fig:TUE} we can
give the physical interpretation of the periodicity of the resonant tunneling
current peaks.
If spacer width $s$ (Fig.~\ref{fig:TLE}) or source-drain voltage $V$ (Fig.~\ref{fig:TUE} )
decrease, quasi-bound state $|n\rangle$ localized in the TQW gains the energy
until it becomes unbound and goes over into the resonance state with
the positive energy.  Figs.~\ref{fig:TLE} and \ref{fig:TUE} demonstrate
that the formation of the resonance states is periodically repeated  when changing $s$ or $V$.

The resonant tunneling condition
in the conventional form (\ref{eq:res_cond}) neglects the formation of
Stark resonances.  If we take these states into account,
condition (\ref{eq:res_cond}) is modified to
\begin{equation}
0 \leq  E_{in} \simeq E_n \simeq E_{CQW}^{\nu} \leq E_F \;,
\label{eq:res_S}
\end{equation}
where $E_n$ is the real part of the energy of the $n$-th Stark resonance.
The upper panels in Figs.~\ref{fig:TLE} and \ref{fig:TUE} show that the
transmission is periodically enhanced by the contributions arising
from the Stark resonance states.
Based on the modified resonant tunneling condition (\ref{eq:res_S})
we propose the following description of the resonant tunneling: the electrons
injected from the source form the Stark resonance states in the TQW in the
left spacer.
If the energy of the Stark resonance is aligned with the energy of the
quasi-bound state in the CQW, the electrons can tunnel through the nanowire with
the transmission coefficient $T \simeq $1.
This means that the resonant tunneling involves the two states: the Stark
resonance in the TQW and the quasi-bound state in the CQW.
The similar two-step tunneling have been studied in our previous
paper,\cite{Wojcik2010} in which we have found the intrinsic oscillations of the
current in the triple-barrier resonant tunneling diode.

The separations, measured on the spacer width scale (Fig.~\ref{fig:TLE})
and on the voltage scale (Fig.~\ref{fig:TUE}), between the subsequent Stark resonances
and also between the corresponding quasi-bound states localized in the TQW
are almost equal to each other.
The strong resonant tunneling peaks appear for the Stark resonances with the
energies slightly larger than zero.

We note that in both the considered cases, i.e., (i) change of the spacer width at the constant
source-drain voltage and (ii) simultaneous change of the source-drain and gate voltages
at the constant spacer width, we are dealing with the changes of the effective electric
field acting in the nanostructure, since in case (i) the change of spacer width $s$
is equivalent to the change of electric field $F = -V/(2s + 2b +w)$.

\begin{figure}
\centering
\includegraphics{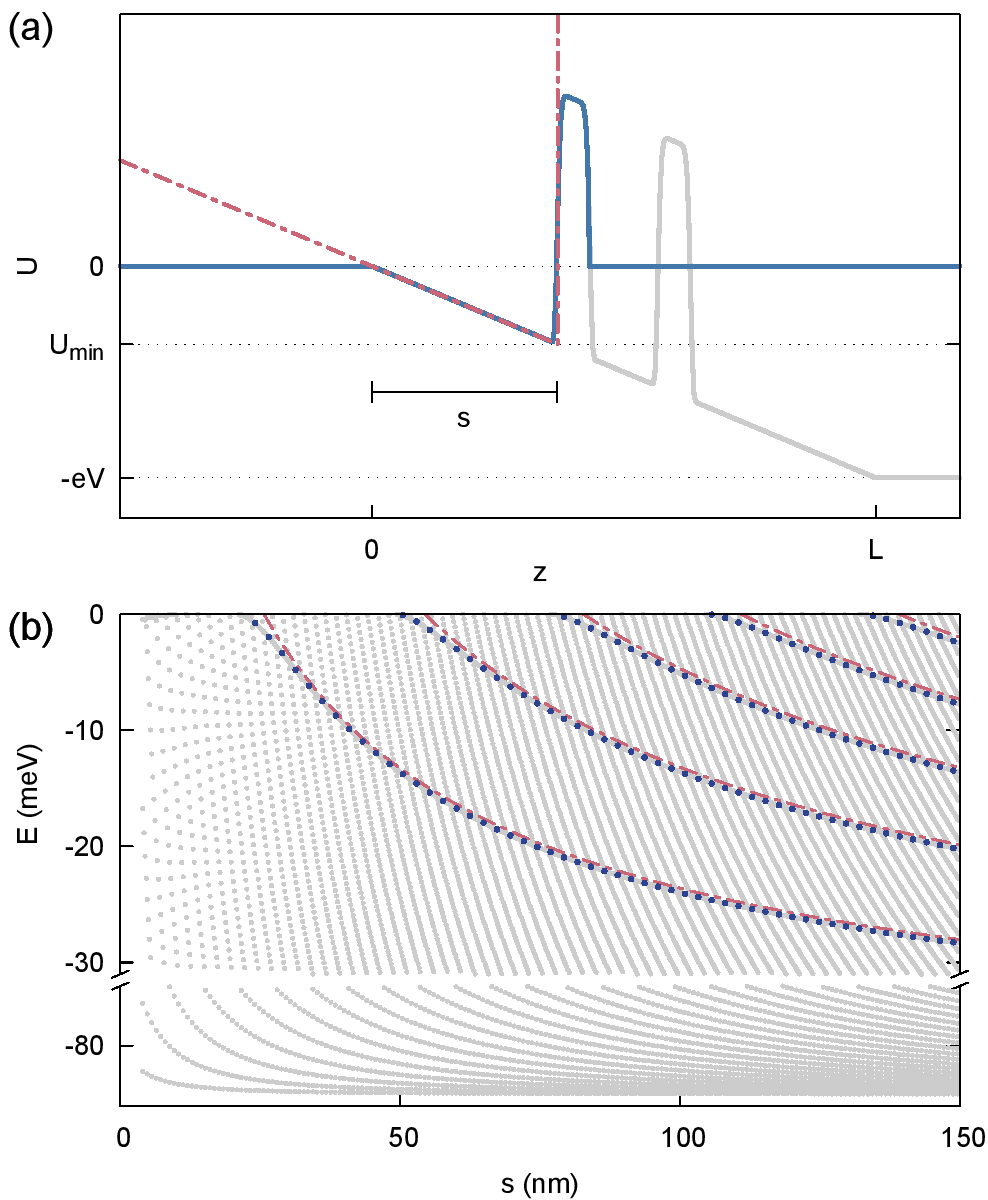} % TQW
\caption{(color online)
(a)~Potential energy profiles for the entire nanosystem (gray lines), the
simple model with the left spacer and barrier (blue lines), and the infinite TQW
(red dash-doted lines) corresponding to the profile of the triangular potential
well in the left spacer region.
The widths of the different regions and the barrier heights are not to scale,
$U_{min}=-eVs/L$.
(b)~Energy eigenvalues calculated for the potential energy profiles plotted in
panel (a) as functions of the spacer width $s$ for $V=84$~mV and $U_G=0$.
The symbols and colors of the curves correspond to those in panel (a).}
\label{fig:TR}
\end{figure}

The previous part of the discussion was based on the numerical results.
In the following part, we will reach the same conclusions based on the
analytical calculations.
For this purpose we propose a simple  model that -- in spite of its simplicity
-- includes the essential physics of the considered nanosystem.
The Stark states are formed in the TQW in the left spacer region.
Therefore,  we construct the simplified potential energy profile (blue curve in
Fig.~\ref{fig:TR}) including the triangular potential and the left potential
barrier only, which accurately approximates the corresponding part of
potential energy plotted in Fig.~\ref{fig:model}.
In order to obtain the analytical results, we make the further approximation
and assume that the triangular potential possesses the infinite depth and range
but the same slope (determined by the electric field $F=-V/L$) as the original
potential energy in the left spacer region (red dash-dotted lines
in Fig. \ref{fig:TR}).
Minimal value $U_{min}$ of the potential energy  (Fig.~\ref{fig:TR})
is determined by width $s$ of the left spacer, i.e., $U_{min} =  - e V s/L$.

In this case, the energy eigenvalues for the infinite triangular potential
can be written as\cite{bookDavies1998}
\begin{equation}
 E_n  =  - \left(  \frac{e^2  \hbar^2 F^2}{2 m}  \right)^{1/3}  \alpha_n
           + U_{min} \;,
  \label{eq:en}
\end{equation}
where $n=1,2,3,\ldots$ and $\alpha_n$ is the $n$-th zero of the Airy function
$\mathrm{Ai}(z)$.
Quantities $\alpha_n$ can be approximated by the
formula\cite{bookAbramowitz1974}
\begin{equation}
 \alpha_n
 \simeq
 -\left[\frac{3\pi}{2}\left(n-\frac{1}{4}\right)\right]^{2/3} \;,
 \label{eq:an}
\end{equation}
which leads to very accurate estimates of the exact values.
The relative errors of estimates (\ref{eq:an}) are very small and decrease
with $n$, taking on the values $0.76\%$, $0.15\%$, $0.06\%$ for $n = 1, 2, 3$,
respectively.\cite{bookAbramowitz1974}

We apply Eqs.~(\ref{eq:en}) and (\ref{eq:an}) to the infinite TQW that
approximates the potential energy of the left spacer with width $s$
and obtain
\begin{equation}
 E_n  =  - \frac{e V s}{L} + \frac{1}{2 m^{1/3}}
           \left[  \frac{3 \pi \hbar e V}{L}
           \left(n-\frac{1}{4}\right) \right]^{2/3} \;,
  \label{eq:en2}
\end{equation}
where $L=2s+2b+w$.
Fig.~\ref{fig:TR}(b) shows that the energy levels calculated from Eq.~(\ref{eq:en2})
for the infinite triangular potential (red dash-dotted lines)
and calculated by the numerical method for the the simple model
with the left spacer and barrier (blue dots) and for the the full potential-energy
profile (gray dots) almost exactly agree with themselves.

Neglecting the small positive shift of the resonance energy (cf. upper parts of
Figs.~\ref{fig:TLE} and \ref{fig:TUE}), we can assume that the maximum of the
resonant tunneling current appears if the Stark state localized in the left
spacer becomes unbound and goes over into the Stark resonance.
Then, the resonant tunneling condition takes on the following approximate form:
\begin{equation}
E_n = 0\;.
\label{eq:E0}
\end{equation}
Using Eqs.~(\ref{eq:en2}) and (\ref{eq:E0}) we have determined the spacer width
$s_n$, for which resonant tunneling condition (\ref{eq:E0}) is satisfied.
This leads to
\begin{equation}
 \frac{2 s_n}{\sqrt{2 + \frac{2b+w}{s_n}}} =
           \frac{3 \pi \hbar}{\sqrt{2 m e V}}
           \left(n-\frac{1}{4}\right) \;.
  \label{eq:sn}
\end{equation}
The numerical solutions of Eq.~(\ref{eq:sn}) for $V=84$~mV and $n=1,2,3$
results in $s_2-s_1=28.3$~nm and $s_3-s_2=27.9$~nm.
These intervals are nearly equal to each other and agree very well with the
differences between the spacer widths that correspond to the subsequent strong
peaks of the resonant tunneling current obtained from the extended numerical
calculations [cf. Fig.~\ref{fig:IL}(b)].

If $s_n \gg 2b+w$, the left hand side of Eq.~(\ref{eq:sn}) can be approximated
by $\sqrt{2}s_n-(b+w/2)/\sqrt{2})$, from which we obtain
\begin{equation}
 s_n = \frac{2b+w}{4} + \frac{3 \pi \hbar}{2 \sqrt{m e V}}
           \left(n-\frac{1}{4}\right) \;.
 \label{eq:limsn}
\end{equation}
The differences between the spacer widths corresponding to the neighboring
strong resonant current peaks are given by
\begin{equation}
 P_s = s_{n+1}-s_{n} =  \frac{3 \pi \hbar}{2 \sqrt{m e V}} \;.
 \label{eq:limdsn}
\end{equation}
Formula (\ref{eq:limdsn}) demonstrates that the separations
between the subsequent resonant current peaks are periodic functions of the
spacer width and are independent of $n$.
For $V=84$~mV, we obtain from Eq.~(\ref{eq:limdsn})
$P_s = 27.6$~nm,
which again is in a good agreement with the estimates obtained from Eq.~(\ref{eq:sn})
and with the results of Fig.~\ref{fig:IL}(b).

Eq.~(\ref{eq:en2}) can also be applied to estimate the values of the source-drain
voltage that correspond to the strong resonant current peaks (Fig.~\ref{fig:IU}).
Using condition (\ref{eq:E0}) we obtain
\begin{equation}
V_{n+1} - V_n =  C \left( n+ \frac{1}{4} \right) \;,
\label{eq:Vn}
\end{equation}
where $C = 9 \pi^2 \hbar^2 L/(4 mes^3)$.
Inserting into Eq.~(\ref{eq:Vn}) the values of the parameters used
in Fig.~\ref{fig:IU}  we get $C = 0.129$~mV and
$V_{n+1} - V_n =$ 4.82, 4.95, 5.08, and 5.20~mV for
$n = 37, 38, 39$, and $40$, respectively.
These values agree very well with the numerical estimates (4.85, 4.99, 5.12, and
5.23~mV) obtained for the same states from Fig.~\ref{fig:IU}(b).
According to Eq.~(\ref{eq:Vn}) the separations
between the subsequent strong resonant current peaks
on the source-drain voltage scale slowly increase with $n$
and can be described by the linear function of $n$.
We remind that $n$ is the quantum number of the Stark state.
Formula (\ref{eq:Vn}) allows us to reproduce the results of the time-consuming
computer simulations in a simple way.
Due to the high accuracy of the results obtained from (\ref{eq:Vn}) this formula
can be used to ascribe the value of quantum number $n$ to the considered state,
which means that we have a tool to identify the Stark resonance state that is responsible
for the given strong resonant current peak.
The analytical results (\ref{eq:limdsn}) and (\ref{eq:Vn}) provide an additional
explanation of the periodic behavior of the resonant tunneling current
presented in Sec.~\ref{sec:result}.

We would like to comment on the use of notation $E_n$ and $E_{CQW}^{\nu}$ for
the one-electron energy levels.
These symbols denote the energy levels of one-electron states
$|n\rangle$ and $|\nu\rangle$, respectively,
that are the eigenstates of the electron with the potential energy depicted in Fig.~\ref{fig:model}.
However, both the numerical and analytical results (\ref{eq:en2}) show that the
states $|n\rangle$ and $|\nu\rangle$ are localized in the two
different parts of the nanostructure, namely, states $|n\rangle$ are localized
in the TQW in the left spacer region and states $|\nu\rangle$ are localized
in the CQW.  Therefore, it is convenient to denote these two subsets of quantum states of the
same system by the different quantum numbers $n$ and $\nu$.

We have also found that the periodic properties of the resonant current peaks
presented in Sec.~\ref{sec:result} can be obtained within the three-dimensional
(3D) model of the nanowire, i.e., can be observed in the realistic nanowires.\cite{Bjork2002}
The results of our preliminary calculations performed with the use of the
adiabatic approximation show that the resonant current peaks are periodic
functions of the spacer width and source-drain voltage.
The present results can be directly applied to the 3D nanowire if zero on the
energy scale is taken at the ground-state energy $E^{gs}_{\perp}$ of the state
that results from the spatial quantization of the lateral electron motion,
i.e., the motion in the $x-y$ plane.
We note that the present 1D model of the electron transport can also be applied
to the mesa-type resonant tunneling structures, for which we can assume the
translational symmetry in the $x-y$ plane and separate the electron motion in
the $z$ direction.

In the semiconductor nanowires, the influence of the impurities on the electron
transport should be taken into account,
since the spacer regions with the sufficiently high purity can hardly be
fabricated.
In the present work, we have neglected the electron-impurity scattering.
Moreover, in the $n$-doped nanowires, the ionization of donors generates the charge that
can be accumulated in the source-related spacer region, which can change the potential profile
and shift the energies of the Stark resonances.
This in turn can disturb the resonant current peaks.
However, in the moderately doped nanowire, due to its small lateral size
only a small fraction of the charge will be gathered in the left spacer, which
will not destroy the periodicity of the resonant current peaks.

\section{Conclusion}\label{sec:concl}

The results of the present paper show that the periodic patterns can be observed
in the resonant tunneling current flowing through the double-barrier structures.
The height of the resonant current peak is a periodic function of the spacer width.
In other words, the resonant tunneling current exhibits periodic changes
if the source-drain distance is changed.  Moreover, the current peaks are periodic
functions of the source-drain voltage if it is changed simultaneously
with the voltage applied to the side gate.
The similar periodicity can appear both in the semiconductor
nanowires and in the resonant tunneling mesa structures.
The physical interpretation of this periodicity is based on the formation
of the Stark states in the triangular quantum well in the spacer region
attached to the source contact.  If the effective electric field becomes weaker,
the quasi-bound Stark states localized in the triangular quantum well cease to be bound
and go over into the resonance Stark states with the energies that enter the
transport window.
This process is periodically repeated when the spacer width and/or source-drain
voltage are changed.  The resonant tunneling current is periodically enhanced if the
electrons tunnel from the source to drain via the Stark resonance state that is
energetically aligned with the quasi-bound state in the CQW.

The Stark resonances, found in the present paper, are the analogues to the
Wannier-Stark resonances predicted for the bulk crystals in external electric
field\cite{Wannier1962} and observed in semiconductor
superlattices.\cite{Morifuji1997}
It is interesting that the infinite triangular potential approximation very
accurately describes the periodicity of the resonant current peaks, which allows
us to explain the physical nature of this effect with the use of a simple
analytically solvable model.

Based on the present results, we can propose a method of experimental
observation of the Stark resonances in semiconductor double-barrier
heterostructures that relies on the measurements of the current-voltage characteristics.
The resonant current peaks that originate from the Stark resonances
should exhibit the periodic behavior as a function of the spacer width,
i.e., source-drain separation, for the constant source-drain voltage
or as functions of source-drain and gate voltages for the nanowires with
the fixed geometric parameters.
We have found that the source-drain voltage difference corresponding to
neighboring strong resonant tunneling current peaks is a linear function of the
Stark state quantum number, which allows us to identify the Stark states in the
double-barrier structures.
We have also shown how to tune the external voltages applied to the
double-barrier structures in order obtain the possibly large differences
between the strong and weak resonant tunneling current peaks,
i.e., to make the current periodicity measurable.

\begin{acknowledgments}
This work has been supported by the National Science Centre, Poland,
under grant No. \mbox{DEC-2011/03/B/ST3/00240}.
\end{acknowledgments}

\end{document}